\documentclass{iopart}
\usepackage{euscript,iopams,amssymb,amsfonts,graphicx,bm}
\usepackage{pgfplots,setstack}
\pgfplotsset{width=6.5cm,compat=1.3}
\usepackage{subcaption}
  \usepackage{paralist}
  \usepackage{epstopdf}
  \usepackage{graphics} 
 \usepackage[colorlinks=true]{hyperref}
 \hypersetup{urlcolor=blue, citecolor=red}
 \usepackage[latin1]{inputenc}

\eqnobysec

\newcommand{\R}{\mathbb{R}}

\newcommand{\x}{\boldsymbol{x}}
\newcommand{\y}{\boldsymbol{y}}

\newcommand{\Markov}[2]{\underset{#1}{\overset{#2}{\rightleftharpoons}}}

\newcommand{\X}{{\bf X}}

\renewcommand{\e}{{\rm e}}

\renewcommand{\P}{\mathbb{P}}

\newcommand{\E}{\mathbb{E}}

\newcommand{\calU}{{\mathcal U}}
\newcommand{\calT}{{\mathcal T}}

\newcommand{\calQ}{{\widetilde Q}}

\begin{document}

\title[Stochastically-gated target]{Diffusive search for a stochastically-gated target with resetting}
\author{Paul C. Bressloff}
\address{Department of Mathematics, University of Utah, Salt Lake City, UT, USA} \ead{bressloff@math.utah.edu}

\begin{abstract}

In this paper, we analyze the mean first passage time (MFPT) for a single Brownian particle to find a stochastically-gated target under the additional condition that the position of the particle is reset to a fixed position $\x_r$ at a rate $r$. The gate switches between an open and closed state according to a two-state Markov chain and can only be detected by the searcher in the open state. One possible example of such a target is a protein switching between different conformational states. As expected, the MFPT with or without resetting is an increasing function of the fraction of time $\rho_0$ that the gate is closed. However, the interplay between stochastic resetting and stochastic gating has non-trivial effects with regards the optimization of the search process under resetting. First, by considering the diffusive search for a gated target at one end of an interval, we show that the fractional change in the MFPT under resetting exhibits a non-monotonic dependence on $\rho_0$. In particular, the percentage reduction of the MFPT at the optimal resetting rate (when it exists) increases with $\rho_0$ up to some critical value, after which it decreases and eventually vanishes. Second, in the case of a spherical target in $\R^d$, the dependence of the MFPT on the spatial dimension $d$ is significantly amplified in the presence of stochastic gating.
\end{abstract}

\maketitle

%%%%%%%%%%%%%%%%%%%%%%%%%%%%%%%%%%%%%%%%%%%%%%%%%%%%%%%%%%%%%%%%%%%%%%%%%%%%%%%%%%%%%%%%%%%
\section{Introduction}

Random search strategies arise throughout nature as a means of efficiently searching for one or more targets of unknown location. Examples include animal foraging
\cite{Bell91,Bartumeus09,Viswanathan11}, proteins searching for particular sites on DNA \cite{Berg81,Halford04,Coppey04,Lange15}, biochemical reaction kinetics \cite{Loverdo08,Benichou10}, and molecular transport within cells \cite{Bressloff13}. Suppose that there is a single fixed target $\calU_0\subset \calU$ in some prescribed domain $\calU\subseteq \R^d$. The searcher is typically represented as a particle whose position $\X(t)$ at time $t$ evolves according to some stochastic process 
\begin{equation*}
\frac{\partial p(\y,t|\x)}{\partial t}={\mathbb L} p(\y,t|\x),\quad \x,\y \in \calU\backslash \calU_0, \ t >0,
\end{equation*}
where $p(\y,t,|\x)$ is the probability density for the particle to be at $\y$ at time $t$ given the initial position $\x$ and
${\mathbb L}$ is the infinitesimal generator of the stochastic process. In the case of diffusive search, ${\mathbb L}=D\nabla^2$, where $D$ is the diffusivity. Target detection is usually implemented by imposing the absorbing boundary condition $p(\y,t|\x)=0$ for all $\y\in \partial \calU_0$, or sometimes a partially absorbing or Robin boundary condition. The efficiency of the search process can then be investigated by solving a corresponding first passage time (FPT) problem. In recent years there have been considerable interest in so-called random intermittent search processes, whereby the particle randomly switches between a slow search phase and a faster non--search phase \cite{Benichou11}, or the position of the particle is reset to a fixed location at a random sequence of times, which is typically (but not necessarily) generated by a Poisson process \cite{Evans20}.

The above basic framework is also the starting point for the classical Smoluchowski theory of diffusion-limited reactions, where the single particle or searcher is replaced by a background sea of particles and the effective flux into the target is identified with the reaction rate \cite{Smoluchowski17,Redner01}. Consider the particular problem of a single stationary protein surrounded by ligands that can bind to the protein. Furthermore, suppose that the target protein can switch between two conformational states $n=0,1$, and is only reactive in the open state $n=1$. That is, the target is stochastically gated. For unbounded domains, this problem was first studied by Szabo {\em et al} \cite{Szabo82}, who assumed that it is irrelevant whether it is the target protein or the diffusing ligands that switch between conformational states. Although the symmetry holds for a pair of reacting particles, it breaks down when a single protein is surrounded by many ligands \cite{Zhou96,Spouge96,Lin98,Benichou00}. In particular, one finds that the kinetics is slower when the gating is due to the protein rather than the ligands, due to the presence of multi-particle correlations in the former case. An analogous problem has been explored within the context of bounded domains \cite{Bressloff15}.

In this paper, we analyze the mean first passage time (MFPT) for a single searcher to find a stochastically-gated target under the additional condition that the position of the particle is reset to a fixed position $\x_r$ at a rate $r$. In contrast to diffusion-limited reactions, the results are independent of whether the searcher or the target is gated. 
Search processes with stochastic resetting have been extensively studied in recent years, as highlighted in the review \cite{Evans20}. A major reason for such interest is that in many cases the MFPT is found to be a unimodal function of the resetting rate $r$ with a unique minimum at an optimal resetting rate $r_{\rm opt}$. Moreover, the optimal search process exhibits certain universal characteristics \cite{Rotbart15,Reuveni16,Pal17,Belan18,Pal20}. One way to determine the MFPT is to exploit the fact that following resetting, the particle has no memory of previous search phases. This means that one can use renewal theory to relate the survival probability with resetting to the corresponding survival probability without resetting, and calculate moments of the FPT density using Laplace transforms. One subtle point is that in the presence of a stochastically-gated target (or searcher), it is necessary to specify a reset rule for the conformational state of the gate. A similar issue arises in the case of other switching processes \cite{Evans18,Bressloff20,Bressloff20a}. 

At first sight, the inclusion of a gated target does not appear particularly interesting from the physical rather than mathematical perspective, since one simply expects the MFPT with or without resetting to increase as the fraction of time $\rho_0$ that the gate is closed increases. However, as we show in this paper, the interplay between stochastic resetting and stochastic gating has non-trivial effects with regards the optimization of the search process under resetting. We first consider an example where the MFPT without resetting, $T_0$, is finite, namely, the diffusive search for a gated target at one end of an interval. We find that the fractional change in the MFPT, $\Delta(r)=T_r/T_0$, where $T_r$ is the MFPT as a function of $r$, exhibits a non-monotonic dependence on $\rho_0$, even though $T_r$ and $T_0$ are monotonically increasing function of $\rho_0$ for fixed $r$. In particular, the percentage reduction of the MFPT at the optimal resetting rate (when it exists) increases with $\rho_0$ up to some critical value, after which it decreases and eventually vanishes. One also finds cases where an optimal resetting rate $r_{\rm opt}$ exists for a range of nonzero values of $\rho_0$ even though $r_{\rm opt}$ does not exist in the absence of a gate. In our second example we consider the search for a spherical target in $\R^d$, which has $T_0=\infty$. Here we show that the MFPT increases monotonically with $\rho_0$ such that the dependence of the MFPT $T_r$ on the spatial dimension $d$ is significantly amplified in the presence of stochastic gating. The structure of the paper is as follows. In section 2 we briefly review how renewal theory can be used to express the Laplace transform of the survival probability with resetting to the corresponding Laplace transform without resetting. In section 3 the general FPT problem for finding a stochastically gated target is formulated. In particular, is is shown how the survival probability without resetting satisfies a boundary value problem (BVP), whose Laplace transform can be used to derive corresponding BVPs for the moments of the FPT density. In sections 4 and 5 we combine the theory presented in the previous two sections to analyze two specific examples, namely a stochastically-gated target on the interval and a stochastically-gated spherical target in $\R^d$.

\section{Search process with stochastic resetting}

\begin{figure}[t!]
\raggedleft
\includegraphics[width=10cm]{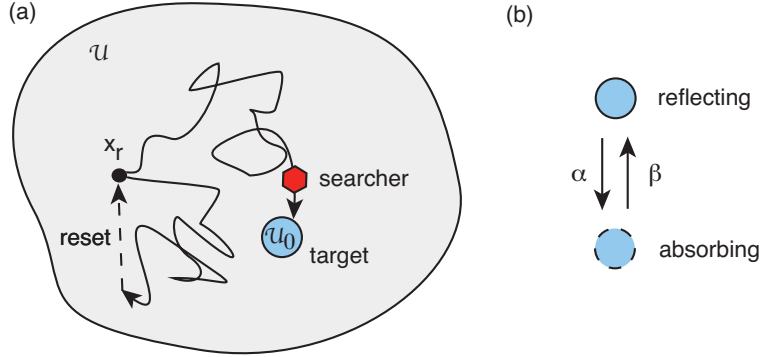}
\caption{(a) Brownian particle searching for a target ${\mathcal U}_{0}$ in a domain $\calU\subseteq \R^d$. Prior to finding the target, the particle can reset to a fixed location $\x_r$ at a rate $r$, after which the search process restarts. (b) The target switches between an absorbing state and a reflecting state according to a two-state Markov chain with transition rates $\alpha,\beta$.}
\label{fig1}
\end{figure}

Consider a Brownian particle (searcher) subject to stochastic motion in a domain $ \calU \subseteq  \R^d$, and resetting to a fixed point $\x_r$ at a rate $r$, see Fig. \ref{fig1}(a). Suppose that there exists some target ${\mathcal U}_{0} \subset \calU$ whose boundary $\partial {\mathcal U}_0 $ is absorbing and $\x_r\notin  {\mathcal U}_0 $. 
The probability density $p_r(\y,t|\x_r)$ for the particle to be at position $\y$ at time $t$  evolves according to the modified diffusion equation 
\begin{equation}
 \label{1}
\frac{\partial p_r(\y,t|\x)}{\partial t}=\nabla_{\y}^2 p_r(\y,t|\x)-rp_r(\y,t|\x)+r\delta(\y-\x_r).
\end{equation}
This is supplemented by the absorbing boundary condition $p_r(\y,t|\x)=0$ for all $\y\in \partial {\mathcal U}_{0} $ and the initial condition $p_r(\y,0|\x)=\delta(\y-\x)$. For simplicity, we also take $\x_r$ to be the initial position of the searcher, $\x=\x_r$. Introduce the survival probability that the particle has not been captured by the target up to time $t$, having started at $\x_r$:
\begin{equation}
Q_r(\x_r,t)=\int_{\calU\backslash {\mathcal U}_{0}}p_r(\y,t|\x_r)d\y.
\end{equation}
Note that $Q_r(\x_r,0)=1$ for $\x_r \notin {\mathcal U}_0$. We assume that for $0<r <\infty$ the particle is eventually captured by the target with probability one so $\lim_{t\rightarrow \infty}Q_r(\x_r,t)=0$.
Let ${\mathcal T}(\x_r)$ denote the first passage time to be absorbed by the target, having started at $\x_r$: 
\[
{\mathcal T}(\x_r)=\inf \{t>0; \ \X(t)\in \partial {\mathcal U}_0 , \, \X(0)=\x_r\}.
\]
The MFPT can be expressed in terms of the survival probability according to \begin{eqnarray}
\label{TQ}
T_r(\x_r)&=\E[{\mathcal T}(\x_r)]=- \int_0^{\infty}t \frac{dQ_r(\x_r,t)}{dt}d\tau=\int_0^{\infty}Q_r(\x_r,t)dt.
\end{eqnarray}
We have used the fact that the FPT density $f_r(t)$ is related to the survival probability according to
\begin{equation}
\label{fr}
f_r(t)=-  \frac{dQ_r(\x_r,t)}{dt}.
\end{equation}

$Q_r$ can be related to the survival probability without resetting, $Q$, using a last renewal equation \cite{Evans11a,Evans11b,Evans20}:
\begin{eqnarray}
\label{renQ}
Q_r(\x_r,t)&=\e^{-rt}Q(\x_r,t) +r\int_0^tQ(\x_r,t')Q_r(\x_r,t-t')\e^{-rt'}dt'. 
\end{eqnarray}
The first term on the right-hand side represents trajectories with no resettings. The integrand in the second term is the contribution from trajectories that last reset at time $t-t'  $, and consists of the product of the survival probability starting from $\x_r$ with resetting up to time $t-t'$ and the survival probability starting from $\x_r$ without any resetting for the time interval of duration $t'$. Since we have a convolution, it is natural to introduce the Laplace transform
\[ \widetilde{Q}_r(\x_r,s)=\int_0^{\infty}Q_r(\x_r,t)\e^{-st}dt.\]
Laplace transforming the last renewal equation and rearranging gives \cite{Evans11a,Evans11b}
\begin{equation}
\label{Qr}
 \widetilde{Q}_r(\x_r,s)=\frac{ \widetilde{Q}(\x_r,r+s)}{1-r \widetilde{Q}(\x_r,r+s)}.
 \end{equation}
Substituting into Eq. (\ref{TQ}) then shows that the MFPT to reach the target is 
\begin{equation}
\label{Tr}
T_r (\x_r)=\widetilde{Q}_r(\x_r,0)=\frac{ \widetilde{Q}(\x_r,r)}{1-r \widetilde{Q}(\x_r,r)}.
 \end{equation}

A major feature of search processes with stochastic resetting is that the MFPT is often a unimodal function of the resetting rate with a minimum at an optimal value $r_{\rm opt}$. One canonical example is diffusion in an unbounded domain $\calU=\R^d$. It is well known that in the absence of resetting ($r=0$) the MFPT is infinite, irrespective of whether diffusion is recurrent or transient. The MFPT is also infinite in the limit $r\rightarrow \infty$, since the particle resets so often that it never reaches the target. One thus finds that the MFPT is minimized at an intermediate value of $r$. One way to investigate whether or not the MFPT has at least one turning point is to calculate the sign of the derivative $dT_r/dr$ at $r=0$ \cite{Reuveni16,Pal17,Belan18,Pal20}. If this derivative is negative then resetting reduces the MFPT in the small-$r$ regime. Equation (\ref{Qr}) implies that
 \begin{eqnarray}
T_r'(\x_r)&=\widetilde{Q}'(\x_r,0)+\widetilde{Q}(\x_r,0)^2=T(\x_r)^2-\frac{T^{(2)}(\x_r)}{2},
\end{eqnarray}
where $T^{(2)}(\x_r)=\E[\calT(\x_r)^2]$ is the second moment of the FPT density. Introducing the variance $\sigma^2(\x_r)=T^{(2)}(\x_r)-T(\x_r)^2$,
 it follows that adding a small rate of resetting reduces the  MFPT for a given $\x_r$ if and only if the coefficient of variation (CV) satisfies
\begin{equation}
\label{cri2}
CV(\x_r):=\frac{\sigma(\x_r)}{  T (\x_r)}>1.
\end{equation}

\section{FPT problem for a stochastically-gated target}

Now suppose that the target is stochastically gated. That is, the boundary $\partial {\mathcal U}_{0}$ of the target switches between an absorbing state $N(t)=1$ and a reflecting state $N(t)=0$ according to a two-state Markov chain:
\begin{equation}
\label{MP}
0\Markov{\beta}{\alpha}1,
\end{equation}
with fixed transition rates $\alpha,\beta$, see Fig. \ref{fig1}(b). 
First consider the case without resetting. Introduce the pair of probability densities
\begin{eqnarray*}
\fl p_n(\y,t|\x,m)d\y=\P[\y<\X(t)<\y+d\y,N(t)=n|\X(0)=\x,N(0)=m].
\end{eqnarray*}
These satisfy the differential Chapman-Kolmogorov (CK) equation
\begin{eqnarray}
\label{CKNa}
\frac{\partial p_{0}}{\partial t}&= D \nabla^2p_{0}-\alpha p_0+\beta p_1,\\
\frac{\partial p_{1}}{\partial t}&= D \nabla^2p_{1}+\alpha p_0-\beta p_1,\quad \y\in \calU\backslash \calU_0,
\label{CKNb}
\end{eqnarray}
where differentiation is with respect to $\y$.
The boundary conditions are
\begin{eqnarray}
\partial_{\sigma} p_{0}(\y,t|\x,m)&=0 ,\  p_1(\y,t|\x,m)=0,  \,  \forall \y \in \partial{\mathcal U}_{0},\nonumber \\
 \partial_{\sigma}p_n(\y,t|\x,m)&=0, \, \forall \y \in \partial \calU,\ n=0,1,
\end{eqnarray}
where $\partial_{\sigma}$ indicates the normal derivative with respect to $\y$, and the initial conditions are
\begin{equation}
p_n(\y,0|\x)=\rho_n\delta(\y-\x),\quad \rho_0=\frac{\beta}{\alpha+\beta},\quad \rho_1=\frac{\alpha}{\alpha+\beta}.
\end{equation}
Here $\rho_m$ is the stationary distribution of the Markov chain.

Let ${\mathcal T}_m(\x)$ denote the FPT given the initial state $(\X(0),N(0))=(\x,m)$ and introduce the survival probabilities
\begin{equation}
S_m(\x,t)=\int_{\calU\backslash {\mathcal U}_{0}}p(\y,t|\x,m)d\y,
\end{equation}
where $p=p_0+p_1$. It can be shown that $S_m$ evolves according to the backward CK equation
\begin{eqnarray}
\label{CKQa}
\frac{\partial S_{0}}{\partial t}&= D \nabla^2S_{0}-\alpha S_0+\alpha S_1,\\
\frac{\partial S_{1}}{\partial t}&= D \nabla^2S_{1}+\beta S_0-\beta S_1,\quad \x\in \calU\backslash \calU_0,
\label{CKQb}
\end{eqnarray}
where differentiation is now with respect to the initial position $\x$.
The boundary conditions are of the same form as the forward equation,
\begin{eqnarray}
\partial_{\sigma} S_{0}(\x,t)&=0 ,\  S_1(\x,t)=0,  \,  \forall \x \in \partial{\mathcal U}_{0},\nonumber\\
 \partial_{\sigma}S_n(\x,t)&=0, \, \forall \x \in \partial \calU,\ n=0,1,
\end{eqnarray}
and the initial conditions are $S_m(\x,0)=1$.
The corresponding FPT densities are given by $f_m(\x,t)=-\partial Q_m(\x,t)/\partial t$ and the $n$-th order moments are
 \begin{eqnarray}
\fl \E[({\mathcal T}_m(\x))^n]&= \int_0t^nf_m(\x,t)dt=\left .\left (-\frac{d}{ds}\right )^n\int_0^{\infty} \e^{-st}f_m(\x,t)dt\right |_{s=0}\\
\fl &=-\left .\left (-\frac{d}{ds}\right )^n\int_0^{\infty} \e^{-st}\frac{\partial S_m}{\partial t}dt\right |_{s=0}=\left .\left (-\frac{d}{ds}\right )^n (1-s\widetilde{S}_m(\x,s))\right |_{s=0},\nonumber 
\end{eqnarray}
where $\widetilde{S}_m(\x,s)$ is the Laplace transform of $S_m(\x,t)$ for $m=0,1$. In particular,
\begin{eqnarray}
T_m(\x)&=\E[{\mathcal T}_m(\x)]=\widetilde{S}_m(\x,0),\nonumber \\ T_m^{(2)}(\x)&=\E[({\mathcal T}_m(\x))^2]=-2\widetilde{S}_m'(\x,0),
\label{TT2}
\end{eqnarray}
where $' $ indicates differentiation with respect to $s$. 

The analysis of resetting in section 2 carries over to the case of a stochastically-gated target, provided that the gate is also reset to the state $m$ with probability $\rho_m$ whenever the particle returns to $\x_r$.\footnote{In the case of a single particle, the search for a stochastically-gated target is equivalent to the search for a non-gated target in which the searcher itself switches between two states; one that reflects off the target and the other that allows absorption by the target. In this scenario it is the state of the searcher that is reset.} The renewal equation (\ref{renQ}) then still holds for $\x=\x_r$ with $\calQ(\x,s)$ the Laplace transform of the weighted survival probability
\begin{equation}
Q(\x,t)=\rho_0S_0(\x,t)+\rho_1S_1(\x,t).
\end{equation}
In order to determine $\calQ$, it is more convenient to work directly with the Laplace transform of the backward equations (\ref{CKQa}) and (\ref{CKQb}). These take the form 
\begin{eqnarray*}
-1&= D \nabla^2\widetilde{S}_0 -(\alpha +s)\widetilde{S}_0+\alpha \widetilde{S}_1,\\
-1&= D \nabla^2\widetilde{S}_{1}+\beta \widetilde{S}_0-(\beta +s)\widetilde{S}_1,\quad \x\in \calU\backslash \calU_0.
\end{eqnarray*}
The boundary conditions are the same as for $S_m$. Now setting 
$\widetilde{Q}_m=\rho_m\widetilde{S}_m$, we have
\begin{eqnarray}
\label{LTQa}
-\rho_0&= D \nabla^2\widetilde{Q}_0 -(\alpha +s)\widetilde{Q}_0+\beta \widetilde{Q}_1,\\
-\rho_1&= D \nabla^2\widetilde{Q}_{1}+\alpha \widetilde{Q}_0-(\beta +s)\widetilde{Q}_1,\quad \x\in \calU\backslash \calU_0.
\label{LTQb}
\end{eqnarray}
An advantage of working with the weighted survival probabilities $\calQ_m$ is that we can add the above pair of equations to obtain a boundary value problem (BVP) involving $\calQ=\calQ_0+\calQ_1$:
\begin{eqnarray}
\label{BVPQa}
&D \nabla^2\calQ-s\calQ=-1,\\
& D \nabla^2 \calQ_{0}-(\alpha+\beta+s) \calQ_0=-\rho_0-\beta \calQ,,\quad \x\in \calU\backslash \calU_0,
\label{BVPQb}
\end{eqnarray}
together with the boundary conditions
\begin{eqnarray}
 \partial_{\sigma}\calQ_0(\x,s)&=0,\  \calQ(\x,s)=\Psi (\x,s),  \,  \forall \x \in \partial{\mathcal U}_{0},\nonumber\\ \partial_{\sigma}\calQ(\x,s)&=0=\partial_{\sigma}\calQ_0(\x,s), \, \forall \x \in \partial \calU.
\end{eqnarray}
The method of solution involves solving the diffusion equation for $W$ in terms of the boundary term $\Psi(\x,s)=\calQ_0(\x,s)$, $\x\in \partial \calU_0$ and then substituting into the inhomogeneous equation for $\calQ_0(\x,s)$. This yields a self-consistency equation for $\Psi(\x,s)$.

The Laplace transforms $\widetilde{Q}_m(\x,s)$ are the generators of the weighted moments. That is,
\begin{eqnarray}
W_m(\x)&:=\rho_mT_m(\x)=\widetilde{Q}_m(\x,0),\nonumber \\Z_m(\x)&:=\rho_mT_m^{(2)}(\x)=-2\widetilde{Q}'_m(\x,0).
\end{eqnarray}
It follows from equations (\ref{BVPQa}) and (\ref{BVPQb}) that $W_0$ and $W=W_0+W_1$ satisfy the BVP
\begin{eqnarray}
\label{BVPTa}
&D \nabla^2W=-1,\\
& D \nabla^2 W_{0}-(\alpha+\beta) W_0=-\rho_0-\beta W,\quad \x\in \calU\backslash \calU_0,
\label{BVPTb}
\end{eqnarray}
together with the boundary conditions
\begin{eqnarray}
\partial_{\sigma} W_0(\x)&=0\ W(\x)=\Phi(\x) ,  \,  \forall \x \in \partial{\mathcal U}_{0}\nonumber \\ \partial_{\sigma}W(\x)&=0=\partial_{\sigma}W_0, \, \forall \x \in \partial \calU,
\end{eqnarray}
with $\Phi(\x)=W_0(\x)$. Similarly, the second moments $Z_0$ and $Z=Z_0+Z_1$ satisfy the BVP
\begin{eqnarray}
\label{BVPT2a}
&D \nabla^2Z=-2W,\\
&D \nabla^2Z_{0} -(\alpha+\beta) Z_0=-2W_0-\beta Z,\quad \x\in \calU\backslash \calU_0,
\label{BVPT2b}
\end{eqnarray}
with $Z=Z_0+Z_1$ and the boundary conditions
\begin{eqnarray}
 \partial_{\sigma} Z_0(\x)&=0\ Z(\x) =\Theta(\x),  \  \forall \x \in \partial{\mathcal U}_{0},\nonumber\\
 \partial_{\sigma}Z(\x)&=0=\partial_{\sigma}Z_0, \, \forall \x \in \partial \calU,
\end{eqnarray}
with $\Theta(\x)=Z_0(\x)$.

\section{Stochastically-gated target on an interval}

 \begin{figure}[t!]
  \raggedleft
    \includegraphics[width=10cm]{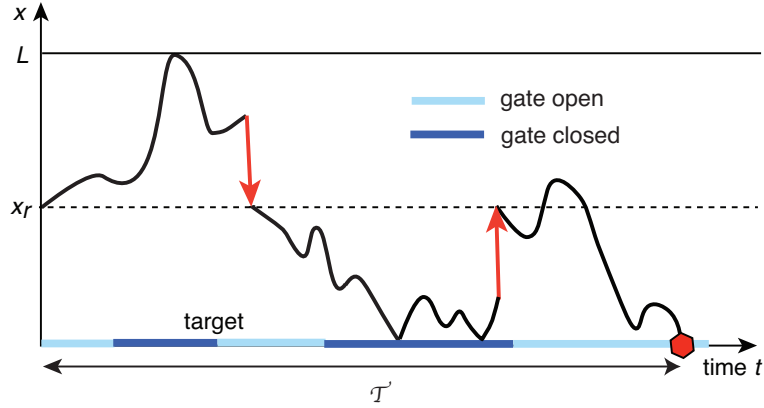}
  \caption{Particle searching for a stochastically-gated target at $x=0$ on the interval $[0,L]$ with a reflecting boundary at $x=L$. After each resetting event, the particle immediately returns to the point $x_r$  and restarts the search phase.}
  \label{fig2}
\end{figure}

We first apply the above theory to the case of diffusion on the interval $[0,L]$ with a stochastically-gated target at $x=0$, see Fig. \ref{fig2}. We proceed by determining the Laplace transformed survival probability without resetting, and then use equation (\ref{Tr}) to investigate the effects of stochastic gating on the MFPT $T_r$.

\subsection{Calculation of the Laplace transform $\calQ$ of the survival probability} 

In the absence of resetting the BVP given by equations (\ref{BVPQa}) and (\ref{BVPQb}) reduces to the 1D form
\begin{eqnarray}
\label{BVPQ1Da}
&D \frac{d^2\calQ}{dx^2}-s\calQ=-1,\\
& D \frac{d^2 \calQ_{0}}{dx^2}-(\alpha+\beta+s) \calQ_0=-\rho_0-\beta \calQ,\quad x\in (0,L),
\label{BVPQ1Db}
\end{eqnarray}
together with the boundary conditions
\begin{eqnarray}
\partial_x\calQ_0(0,s)&=0,\  \calQ(0,s)=\Psi(s),  \nonumber\\ \partial_{x}\calQ(L,s)&=0=\partial_{x}\calQ_0(L,s).
\end{eqnarray}
Equation (\ref{BVPQ1Da}) has the solution
\begin{equation}
\label{solQ}
\calQ(x,s)=\frac{1}{s}+\left (\Psi(s)-\frac{1}{s}\right )\frac{\cosh(\sqrt{s/D}[L-x])}{\cosh(\sqrt{s/D}L)}.
\end{equation}
Equation (\ref{BVPQ1Db}) can be solved in terms of the Neumann Greens function $G_s(x,x')$ where
\begin{equation}
\label{calGs}
\fl  \frac{d^2G_s}{dx^2}-\frac{(\alpha+\beta+s)}{D} G_s=-\delta(x-x'); \ \frac{dG_s}{dx}(0,x')=0; \ \frac{dG_s}{dx}(L,x')=0.
 \end{equation}
 From the divergence theorem
 \begin{equation}
 \int_0^LG_s(x,x')dx=\frac{D}{\alpha+\beta+s}.
 \end{equation}
 We find that
 \begin{equation}
 \label{calG0}
\fl  G(x,x';s)=\frac{1}{\gamma_s \sinh(\gamma_s L)}\left \{ \begin{array}{cc}\cosh(\gamma_s x)\cosh(\gamma_s(L-x')),& x<x'\\
\cosh(\gamma_s [L-x]) \cosh(\gamma_s x'),&x>x'\end{array}\right . ,
\end{equation}
where $\gamma_s = \sqrt{(\alpha+\beta+s)/D}$. The corresponding solution for $\calQ_0(x,s)$ is
\begin{equation}
\calQ_0(x,s)=\frac{1}{D}\int_0^LG(x,x';s)[\rho_0+\beta \calQ(x',s)]dx' .
\end{equation}
Setting $\calQ_0(0,s)=\Psi(s)$ and substituting for $\calQ(x,s)$ yields
\begin{eqnarray}
\Psi(s)&=\frac{1}{(\alpha+\beta+s)}\left [\frac{\beta}{\alpha+\beta}+\frac{\beta}{s }\right ]\\
&\quad +\frac{\beta}{D} \left (\Psi(s)-\frac{1}{s}\right )\int_{0}^LG(0,y;s) \frac{\cosh(\sqrt{s/D}[L-y])}{\cosh(\sqrt{s/D}L)} dy.\nonumber
\end{eqnarray}
Substituting for $G(0,y;s)$ and evaluating the integrals, we obtain the result
\begin{eqnarray}
\fl \Psi(s)&=\frac{1}{(\alpha+\beta+s)}\left [\frac{\beta}{\alpha+\beta}+\frac{\beta}{s}\right ] +\frac{\beta}{D} \left (\Psi(s)-\frac{1}{s}\right )\frac{R(s)}{2\gamma_s\sinh(\gamma_s)},
\end{eqnarray}
with
\begin{eqnarray}
R(s)&=\frac{1}{\cosh(\sqrt{s})}\left [\frac{\sinh(\gamma_s+\sqrt{s/D})}{\gamma_s+\sqrt{s/D}}+\frac{\sinh(\gamma_s-\sqrt{s/D})}{\gamma_s-\sqrt{s/D}}\right ].
\end{eqnarray}
Finally, rearranging gives
\begin{eqnarray}
\fl \Psi(s)&=\left [\frac{1}{(\alpha+\beta+s)}\left (\frac{\beta}{\alpha+\beta}+\frac{\beta}{s}\right ) -\frac{\beta R(s)}{2sD\gamma_s\sinh(\gamma_s)}\right ]\left [1-\frac{\beta R(s)}{2D\gamma_s\sinh(\gamma_s)}\right ]^{-1}.
\end{eqnarray}

Given $\calQ(x,s)$ we can now generate the MFPT $T$ and higher moments, although  taking the limit $s\rightarrow 0$ requires an application of L'Hopitals rule. For example, in the case of a permanently open gate ($\beta=0$), we recover the classical results
\begin{equation}
\label{clas}
\fl T=\tau(x):=-\frac{x^2}{2D}+\frac{xL}{D},\quad T^{(2)}=\tau^{(2)}(x):=\frac{1}{12D^2}(x^4-4x^3L+8xL^3).
\end{equation}

\subsection{Calculation of MFPT $T$} 
A useful check of the analysis is to derive the weighted MFPT $W(x)$ by solving the 1D version of the BVP given by equations (\ref{BVPTa}) and (\ref{BVPTb}) :
\begin{eqnarray}
\label{BVPT1Da}
&D \frac{d^2W}{dx^2}=-1,\\
& D\frac{d^2W_{0}}{dx^2}-(\alpha+\beta) W_0=-\rho_0-\beta W,\quad x\in (0,L),
 \label{BVPT1Db}
\end{eqnarray}
together with the boundary conditions $\partial_xW_0(0)=0,\ W(0)=\Phi$ and $\partial_xW_0(L)=\partial_xW(L)=0$.
Equation (\ref{BVPT1Da}) has the solution
\begin{equation}
\label{solW0}
W(x)=-\frac{x^2}{2D}+\frac{xL}{D}+\Phi=\tau(x)+\Phi,
\end{equation}
where $\tau(x)$ is the MFPT without stochastic gating.
Equation (\ref{BVPT1Db}) can be solved in terms of the Neumann Greens function $G_0(x,x')=G(x,x';0)$, which is obtained by setting $s=0$ in equation (\ref{calG0}). The corresponding solution for $W_0(x)$ is
\begin{equation}
W_0(x)=\frac{1}{D}\int_0^LG_0(x,x')[\rho_0+\beta W(x')]dx' .
\end{equation}
Setting $W_0(0)=\Phi$, substituting for $W(x)$, and rearranging yields
\begin{eqnarray*}
\Phi&=\frac{\beta}{\alpha(\alpha+\beta)}+\frac{\beta(\alpha+\beta)}{\alpha D} \int_{0}^LG_0(0,y)\left [-\frac{y^2}{2D}+\frac{yL}{D}\right ]dy\\
&=\frac{\beta}{\alpha(\alpha+\beta)}+\frac{\beta(\alpha+\beta)}{\alpha D^2\gamma_0 \sinh(\gamma_0 L)}\int_0^L\cosh([L-y]\gamma_0)\left [-\frac{y^2}{2}+{y L}\right ]dy.
\end{eqnarray*}
Set $\tau=L^2/D$ and fix the units of length by taking $L=1$. Then
\begin{eqnarray*}
\Phi=\frac{\beta}{\alpha(\alpha+\beta)}+\frac{\beta(\alpha+\beta)\tau^2}{\alpha \gamma_0 \sinh(\gamma_0 )}\int_0^1\cosh([1-y]\gamma_0)\left [-\frac{y^2}{2}+{y }\right ]dy.
\end{eqnarray*}
Evaluating the integrals by noting that
\[\int_0^1\e^{-\gamma_0 y}y^ndy=\left (-\frac{d}{d\gamma_0}\right )^n\int_0^1\e^{-\gamma_0 y}dy=\left (-\frac{d}{d\gamma_0}\right )^n\left (\frac{1-\e^{-\gamma_0 }}{\gamma_0} \right ),
\]
we find that
\begin{eqnarray}
\Phi=\frac{\beta}{\alpha(\alpha+\beta)}+\frac{\beta(\alpha+\beta)\tau^2}{\alpha \gamma_0^4 \sinh(\gamma_0 )}\left [\gamma_0\cosh(\gamma_0)-\sinh(\gamma_0)\right ].
\end{eqnarray}
It can be checked that the solution (\ref{solW0}) is obtained by taking the limit $\lim_{s\rightarrow 0} \calQ(x,s)$ with $\calQ$ given by equation (\ref{solQ}).

\begin{figure}[t!]
  \raggedleft
  \includegraphics[width=8cm]{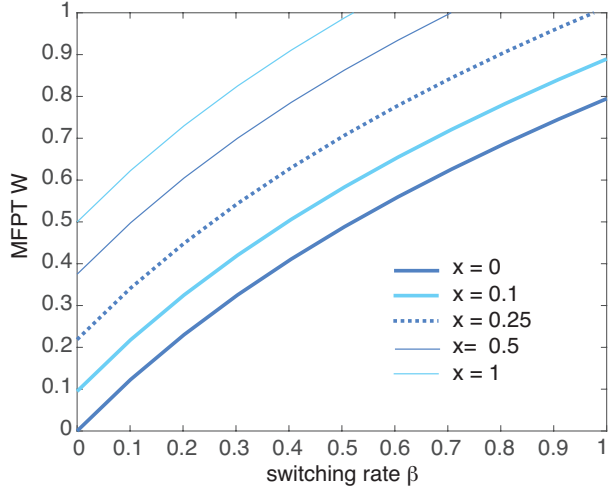}
  \caption{Plot of MFPT $W$ (no resetting) as a function of the switching rate $\beta$ for various initial positions $x$.}
  \label{fig3}
\end{figure}

Let us also fix the time-scale by setting $\alpha =1$ and noting that $\beta=0$ corresponds to a permanently open gate whereas $\beta\rightarrow \infty$ represents the limit in which the gate is always closed. As one would expect, the MFPT $W(x)$ is an increasing function of $\beta$ for all $x\in [0,1]$, see Fig. \ref{fig3}. However, as we show below, gating has non-trivial affects on the optimization of the MFPT in the presence of resetting.

 \begin{figure}[t!]
 \raggedleft
  \includegraphics[width=8cm]{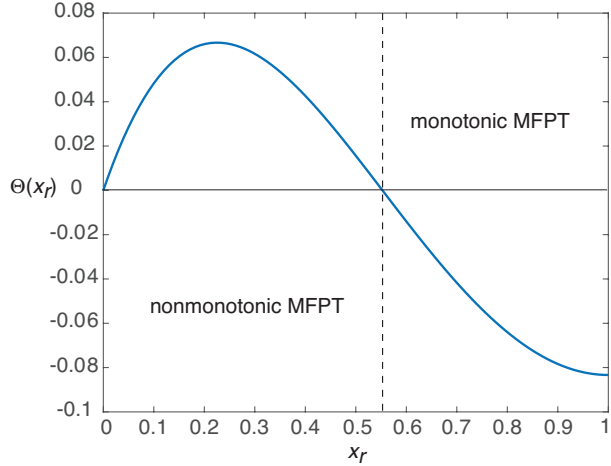}
  \caption{Plot of $\Theta(x_r):=\tau^{(2)}(x_r)-2\tau(x_r)^2$ as a function of the reset location $x_r$. Here $\tau$ and $\tau^{(2)}$ are the MFPT and second moment in the absence of resetting and a stochastic gate, see equation (\ref{clas}). The sign of $\Theta$ determines whether or not the corresponding MFPT with resetting is initially a decreasing function of the resetting rate $r$, see equation (\ref{cri2}).}
  \label{fig4}
\end{figure}

 \subsection{Results for MFPT with resetting, $T_r(x_r)$}
 
We now wish to explore the combined effects of stochastic resetting ($r>0$) and stochastic switching ($\beta >0$). Substituting the solution (\ref{solQ}) into equation (\ref{Tr}) with $x=x_r$ determines the MFPT $T_r(x_r)$ as a function of both $\beta$ and $r$:
\begin{equation}
\label{Tr1D}
\fl T_r (x_r)=\widetilde{Q}_r(x_r,0)=\frac{ r^{-1}\cosh(\sqrt{r/D}L)+\left (\Psi(r)-r^{-1}\right ) \cosh(\sqrt{r/D}[L-x])}{\left (r\Psi(r)-1\right ) \cosh(\sqrt{r/D}[L-x])}.
 \end{equation}
It can be checked that
\[T_0(x_r)=\lim_{r\rightarrow 0}T_r(x_r)=W(x_r).\]
In the absence of gating ($\beta=0$), one finds that the MFPT $T_r(x_r)$ is a unimodal function of $r$ for reset locations close to the target and a monotonically increasing function of $r$ at more distal locations, see also \cite{Ray19,Ray20}; in the former case there exists an optimal resetting rate that minimizes $T_r$. This result can also be understood by plotting $\Theta(x_r):=\tau^{(2)}(x_r)-2\tau(x_r)^2$ as a function of  $x_r$, with $\tau,\tau^{(2)}$ defined in equation (\ref{clas}). Applying the condition (\ref{cri2}) implies that the sign of $\Theta$ determines whether or not the corresponding MFPT $T_r$ with resetting is initially a decreasing function of the resetting rate $r$. It can be seen from Fig. \ref{fig4} that there exists a critical location $x_c\approx 0.55$ such that $\Theta(x_r)$ is negative in the case of proximal positions $x_r<x_c$ but switches to positive values in the case of distal locations $x_r>x_c$. This suggests that when the gate is always open, $T_r$ will be a monotonically increasing function of $r$ for distal locations (eg. $x_r=1$) and a unimodal function of $r$ for proximal locations (eg. $x_r=0.5$). How does the introduction of a stochastic gate affect the optimal search process when it exists?

 \begin{figure}[t!]
  \raggedleft
  \includegraphics[width=10cm]{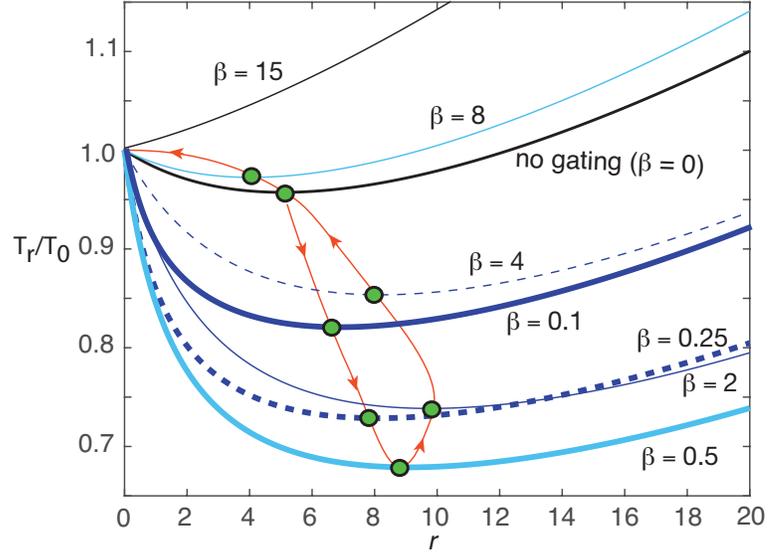}
  \caption{Plot of $\Delta_r=T_r(x_r)/T_0(x_r)$ as a function of the resetting rate $r$ for $x_r=0.5$ and various switching rates $\beta$. Other parameters are $L=1=D$ and $\alpha=1$. The filled green dots indicate the optimal resetting rate for a given $\beta$.}
  \label{fig5}
\end{figure}

\begin{figure}[t!]
  \raggedleft
  \includegraphics[width=10cm]{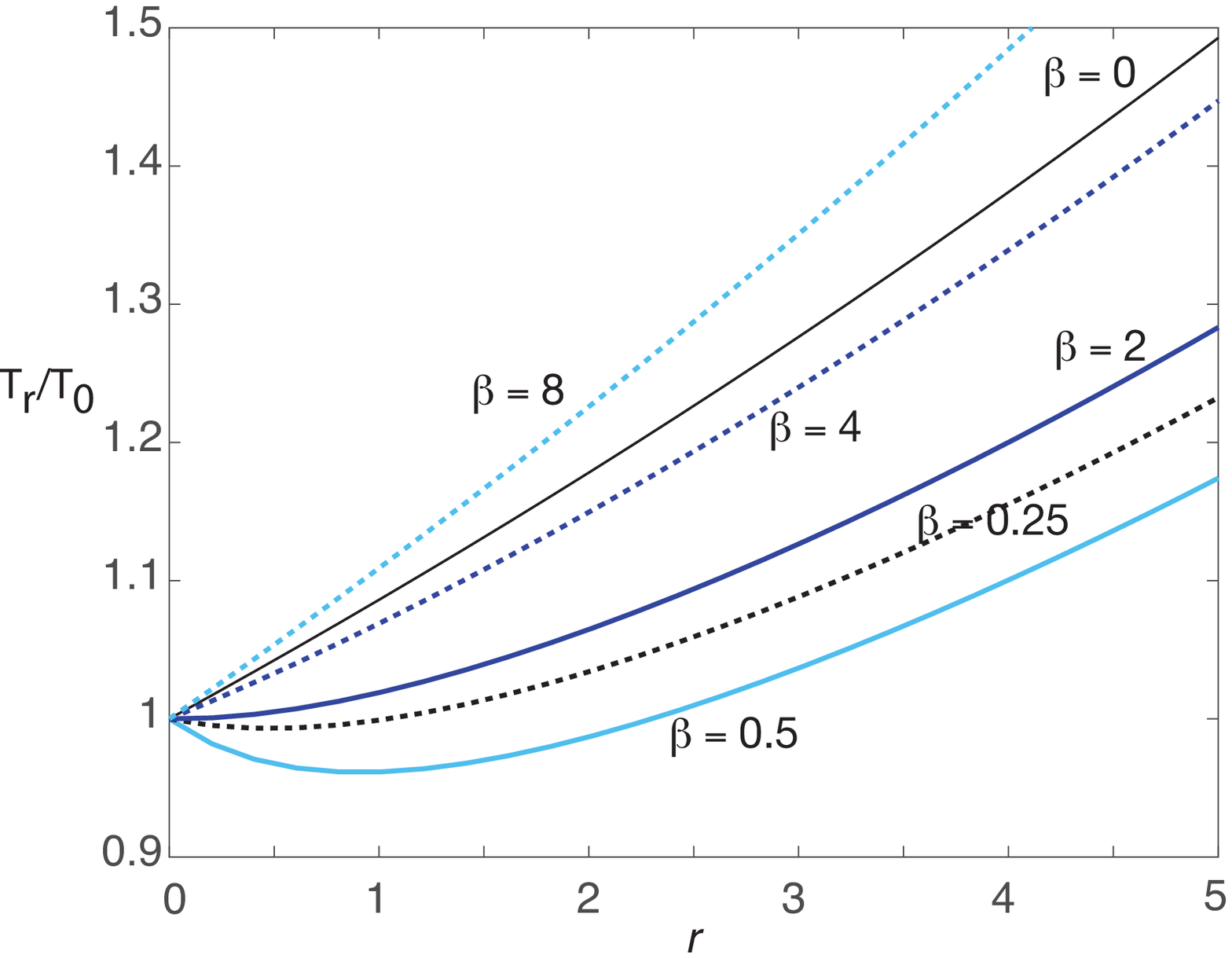}
  \caption{Same as Fig. \ref{fig5} except that $x_r=1$.}
  \label{fig6}
\end{figure}

\begin{figure}[t!]
  \raggedleft
  \includegraphics[width=10cm]{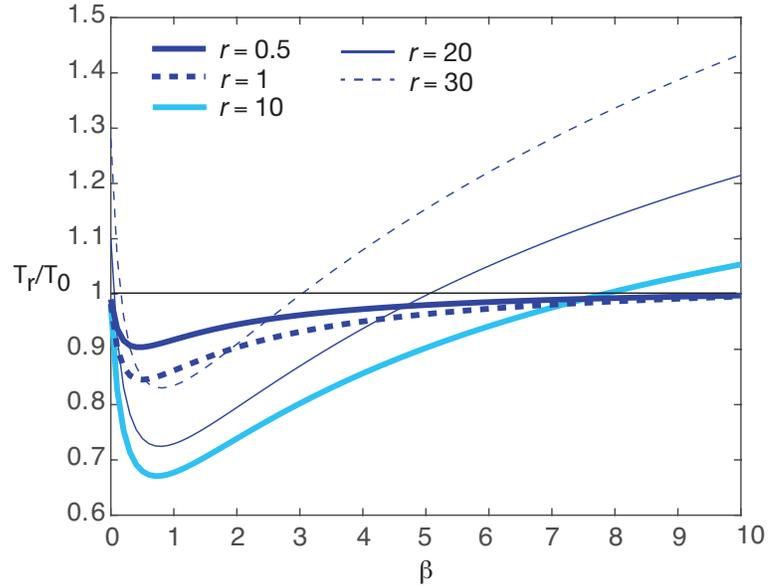}
  \caption{Plot of $\Delta_r=T_r(x_r)/T_0(x_r)$ as a function of the switching rate $\beta$ for $x_r=0.5$ and various resetting rates $r$. Other parameters are as in Fig. \ref{fig5}.}
  \label{fig7}
\end{figure}

\begin{figure}[t!]
\raggedleft
  \includegraphics[width=10cm]{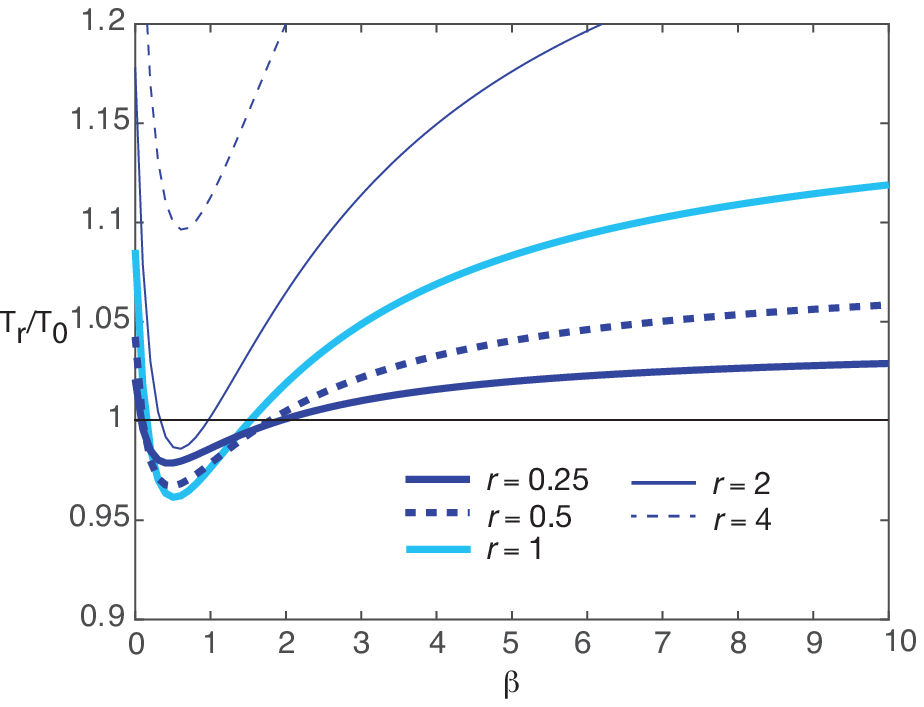}
  \caption{Same as Fig. \ref{fig7} except that $x_r=1$.}
  \label{fig8}
\end{figure}

In Fig. \ref{fig5} we show plots of the normalized MFPT $\Delta_r=T_r(x_r)/T_0(x_r)$ as a function of the resetting rate $r$ for  $x_r=0.5$ and various switching rates $\beta$. Note that for $\beta=0$ the MFPT is a unimodal function of $r$, as expected from Fig. \ref{fig4}. The plots also yield an unexpected result, namely, that in the presence of resetting, the fractional MFPT $\Delta_r$ exhibits a non-monotonic dependence on the switching rate $\beta$, even though both $T_r$ and $T_0$ are monotonically increasing functions of $\beta$. In particular, the percentage reduction of the MFPT at the optimal resetting rate increases with $\beta$ up to some critical value $\beta_c$, after which it decreases again. For sufficiently large $\beta$, resetting increases the MFPT for all $r$. The non-monotonic dependence on $\beta$ is also demonstrated by the functional dependence of the optimal resetting rate $r_{\rm opt}(\beta)$. Stochastic gating also has a significant effect on distal locations $x_r$ for which $T_r$ is a monotonic function of $r$ when $\beta=0$. This is illustrated in Fig. \ref{fig7}, which demonstrates how $T_r$ can become unimodal at intermediate values of $\beta$. In Fig. \ref{fig7} we show the corresponding variation of $\Delta_r$ with respect to $\beta$ for different resetting rates $r$. The non-monotonic $\beta$-dependence is clearly seen. 
The parameter region within which resetting leads to a reduction in the MFPT is diminished for more distal reset locations, as illustrated in Fig. \ref{fig8} for $x_r=1$.
Finally, note that these results also imply that the critical location $x_rx_c$ where $T_r(x_r)$ switches from unimodal to monotonic behavior alos shifts with $\beta$.

\section{Stochastically-gated spherical target}

As our second example, we consider a Brownian particle searching for a $d$-dimensional, stochastically-gated, spherical target in an unbounded domain. The corresponding problem for an absorbing target was analyzed in \cite{Evans14}. In both cases one can exploit spherical symmetry by taking the center of the target to be at the origin so that the solution for the survival probability and moments of the FPT density only depend on the radial distance $x=|\x|$ of the initial position/reset point and the radius $a$ of the sphere. The Neumann boundary conditions are replaced by the far-field condition that solutions remain finite as $|\x|\rightarrow \infty$. In contrast to the previous 1D example, the MFPT is infinite in the absence of resetting.
\subsection{Calculation of $\calQ$}

The BVP given by equations (\ref{BVPQa}) and (\ref{BVPQb}) become
\begin{eqnarray}
\label{BVPRa}
&\frac{d^2\calQ}{dx^2}+\frac{d-1}{x}\frac{d\calQ}{dx}-s\calQ=-1,\\
& \frac{d^2\calQ_0}{dx^2}+\frac{d-1}{x}\frac{d\calQ_0}{dx}-(\alpha+\beta+s) \calQ_0=-\rho_0-\beta \calQ, \quad a<x<\infty,
\label{BVPRb}
\end{eqnarray}
together with the boundary conditions
\begin{eqnarray}
 \partial_{x}\calQ_0(a,s)&=0,\  \calQ(a,s)=\Psi (s).
\end{eqnarray}
Following \cite{Evans14}, the solution of equation (\ref{BVPRa}) takes the form
\begin{equation}
\label{solQ2d}
\calQ(x,s)=\frac{1}{s}-A(s)x^{\nu}K_{\nu}(\eta_sx),
\end{equation}
where $\nu=1-d/2$ and $\eta_s=\sqrt{s/D}$,
and $K_{\nu}$ is the modified Bessel function of the second kind of order $\nu$. The boundary condition $\calQ(a,s)=\Psi(s)$ then relates $A(s)$ to $\Psi(s)$ according to
\begin{equation}
\label{Psi2d}
\Psi(s)=\frac{1}{s}-a^{\nu}K_{\nu}(\eta_s a)A(s).
\end{equation}

In order to determine $\Psi(s)$ and hence $A(s)$, we have to solve equation (\ref{BVPRb}) for $\calQ_0(x,s)$, which can be expanded as
\begin{equation}
\fl \calQ_0(x,s)=C(s)x^{\nu}K_{\nu}(\gamma_sx)+a(s)\calQ(x,s)+b(s),\quad \gamma_s=\sqrt{\frac{s+\alpha+\beta}{D}}
\end{equation}
Substituting back into equation (\ref{BVPRb}) and using (\ref{BVPRa}), we find that
\[-(\alpha+\beta)a(s)\calQ(x,s)-a(s)-(\alpha+\beta+s)b(s)=-\rho_0-\beta \calQ(x,s).
\]
Hence, $a(s)=\rho_0$ and $ b(s)=0$.
Finally, the coefficient $C(s)$ is obtained by imposing the Neumann boundary condition on the surface $r=a$:
\[\fl C(s)a^{\nu}\gamma_s[K_{\nu}'(\gamma_sa)+\nu K_{\nu}(\gamma_sa)/a\gamma_s]-\rho_0a^{\nu}\eta_sA(s)[K_{\nu}'(\eta_sa)+\nu K_{\nu}(\eta_sa)/a\eta_s]=0.\]
Using the identity
\begin{equation*}
K_{\nu}'(x)=-\frac{\nu}{x}K_{\nu}(x)-K_{\nu-1}(x),
\end{equation*}
we have
\[C(s)=\frac{\rho_0 \eta_s K_{\nu-1}(\eta_sa)}{\gamma_s K_{\nu-1}(\gamma_sa)} A(s).
\]
Combining these various results leads to the solution
\begin{eqnarray}
\calQ_0(x,s)= \frac{\rho_0}{s}-\rho_0 x^{\nu}\left [K_{\nu}(\eta_sx)-\frac{\eta_s K_{\nu-1}'(\eta_s a)}{ \gamma_s K_{\nu-1}'(\gamma_sa)}K_{\nu}(\gamma_sx)\right ]A(s).
\end{eqnarray}
Now setting $x=a$ and using equation (\ref{Psi2d}) yields an implicit equation for $A(s)$:
\begin{equation}
\fl \frac{1}{s}-a^{\nu}K_{\nu}(\eta_s a)A(s)=\frac{\rho_0}{s}+\rho_0 a^{\nu}\left [K_{\nu}(\eta_sa)-\frac{\eta_s K_{\nu-1}'(\eta_s a)}{ \gamma_s K_{\nu-1}'(\gamma_sa)}K_{\nu}(\gamma_sa)\right ]A(s),
\end{equation}
which on rearranging gives
\begin{eqnarray}
\label{solA}
A(s)=\frac{\rho_1}{sa^{\nu}}\left [\rho_1 K_{\nu}(\eta_sa)+\rho_0\frac{\eta_s K_{\nu-1}'(\eta_s a)}{ \gamma_s K_{\nu-1}'(\gamma_sa)}K_{\nu}(\gamma_sa)
\right ]^{-1}.
\end{eqnarray}

\begin{figure}[t!]
  \raggedleft
  \includegraphics[width=10cm]{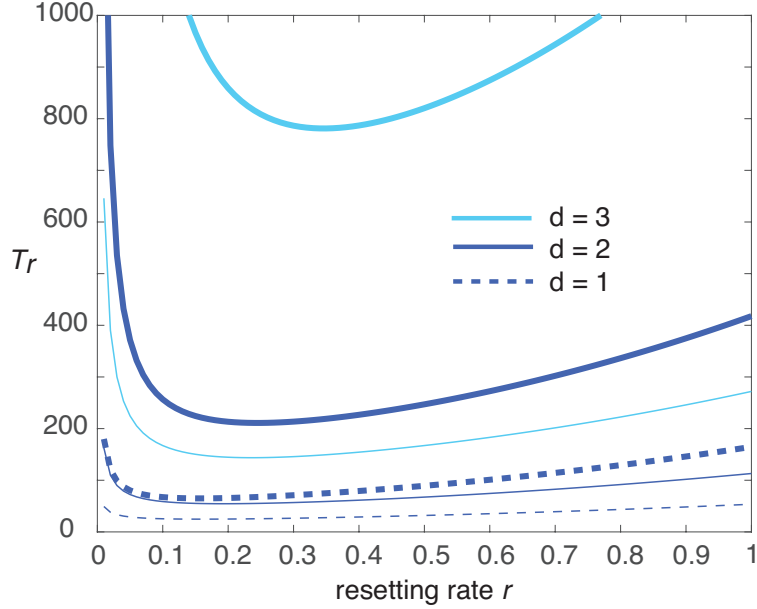}
  \caption{Effect of spatial dimension $d$ on MFPT for a spherical target. Plot of MFPT $T_r(x_r)$ as a function of the resetting rate $r$ for $x_r=5$ and $d=1,2,3$. The three thick curves are for $\beta=4$ and  the three thin curves are for $\beta=0$ (no stochastic gating). Other parameters are $a=1=D$ and $\alpha=1$.}
  \label{fig9}
\end{figure}

\subsection{Results for MFPT with resetting ($T_r$)}
Substituting equation (\ref{solQ2d}) into (\ref{Tr}) with $x=x_r$ leads to the following expression for the MFPT $T_r$ in the presence of resetting:
\begin{equation}
T_r(x_r)=\frac{r^{-1}-A(r)x_r^{\nu}K_{\nu}(\sqrt{r/D}x_r)}{rA(r)x_r^{\nu}K_{\nu}(\sqrt{r/D}x_r)},
\end{equation}
with $A(r)$ given by equation (\ref{solA}). Note that in the limit $\beta\rightarrow 0$ we have $\rho_1\rightarrow 1$ and $\rho_0\rightarrow 0$, and we recover the result of \cite{Evans14}:
\begin{equation}
T_r(x_r)=\frac{a^{\nu}K_{\nu}(\sqrt{r/D}a)- x_r^{\nu}K_{\nu}(\sqrt{r/D}x_r)}{r x_r^{\nu}K_{\nu}(\sqrt{r/D}x_r)}.
\end{equation}
In Fig. \ref{fig9} we show plots of $T_r$ as a function of $r$ for $d=1,2,3$ and various switching rates $\beta$. As expected, for a each value of $\beta$ and $d$, the MFPT is a unimodal function of $r$ with a minimum at some optimal resetting rate. More significantly, $T_r$ is more sensitive to the spatial dimension for larger values of $\beta$.

\section{Discussion}

In this paper we have shown how the interplay between stochastic resetting and the stochastic gating of a target can lead to non-trivial effects with regards the optimization of the search process under resetting. First, in the case of diffusive search on the interval for which the MFPT without resetting is finite, we showed that resetting can be more effective in the presence of gating. That is, the fractional reduction in the MFPT is amplified over an intermediate range of values of the switching rate $\beta$ (or equivalently the fraction of time $\rho_0$ that the gate is closed). Moreover, one finds cases where an optimal resetting rate $r_{\rm opt}(\beta)$ exists for a range of $\beta$-values even though $r_{\rm opt}(0)$ does not exist. A second non-trivial consequence of gating occurs in cases where the MFPT without resetting is infinite, such as the diffusive search for a spherical target in $\R^d$.  In this case, resetting becomes more sensitive to the spatial dimension of the underlying search process when stochastic gating is included. Given these observations, it would be interesting to explore how robust these effects are with respect to the particular choice of the generator ${\mathbb L}$ of the search process, and the inclusion of delays such as refractory periods \cite{Evans19a,Mendez19a} and finite return times \cite{Mendez19,Bodrova20,Pal20,Bressloff20b}. Another extension would be to consider multiple targets and/or multiple searchers; in the latter case one would need to take into account multi-particle correlations when the targets rather than searchers are gated.

\end{document}